\documentclass[nofootinbib,aps,12pt]{revtex4-1}
\usepackage[top=0.8in, bottom=0.9in, left=1.22in, right=1.22in]{geometry}

\begin{document}

\title{Higgs Multiplets of the Quark-Lepton Family Group}
\author{{\bf M.A. Ajaib and S.M. Barr} \\
Bartol Research Institute \\ University of Delaware \\
Newark, Delaware 19716} 

\begin{abstract}
It is shown that realistic models can be constructed in which the
Standard Model Higgs field is in a non-trivial multiplet of a
non-abelian family group of the quarks and leptons. It is shown that
the observed quark and lepton masses and mixing angles can be fit,
while the coefficients of flavor-changing four-fermion operators
mediated by the extra Higgs doublets are determined in terms of only
a few unknown parameters.
\end{abstract} \maketitle

\section{Introduction}

\vspace{0.5cm}
A peculiar feature of the Standard Model is that there are many
multiplets of fermions, but only one multiplet of spin-0 bosons, the
Higgs doublet. Supersymmetrizing the Standard Model would produce a
balance between spin-0 and spin-1/2, but still wouldn't explain why
there are so many matter multiplets (i.e. quarks and leptons) and so
few Higgs multiplets.

\vspace{0.2cm}
In this paper we pursue a different idea than supersymmetry. We
suppose that there is a non-abelian family group \cite{family} under
which {\it both} the Higgs fields and the matter fields transform as
non-trivial multiplets. The particular model we shall describe as an
example has an $SO(4)_F$ family group, under which four quark and
lepton families transform as a 4-plet, a mirror family transforms as
a singlet, and nine Higgs doublets transform as a 9-plet (i.e. as a
rank-2, symmetric, traceless tensor). We shall call all nine of
these doublets ``Higgs" doublets, even though only the lightest of
them --- the Standard Model Higgs doublet --- actually gets a
non-zero vacuum expectation value.

\vspace{0.2cm}
Such a rich Higgs sector would yield new physics beyond the Standard
Model. Most obviously, it would imply the existence of
flavor-changing couplings of the ``extra" Higgs doublets. The
non-abelian family group, besides explaining to some extent why
there are families of quarks and leptons, and giving a rich Higgs
sector, would also greatly constrain the form of the quark and
lepton mass matrices and the couplings of the extra Higgs doublets.
There is therefore the potential of great predictivity. For example,
in the illustrative $SO(4)_F$ model discussed in this paper we shall
show that there are sufficiently many model parameters to give a
good fit to the quark and lepton masses and mixings, but still few
enough parameters that the coefficients of all the flavor-changing
four-fermion operators are almost completely determined.

\vspace{0.2cm}
One might worry that these flavor-changing effects would be too
large. However, in the kind of model we are describing there is a
mass hierarchy within the family multiplet of Higgs fields that
mirrors the mass hierarchy among the families of quarks and leptons.
Therefore, most of the extra Higgs doublets (particularly those that
couple most strongly to the first family of quarks and leptons) are
much heavier than the Standard Model Higgs doublet, and excessive
flavor-changing effects can be avoided. Nevertheless, as will be
seen, there typically is a ``lightest extra Higgs doublet" (LED)
that can give flavor-changing near the current limits.

This raises another question: given that there is no low-energy
supersymmetry to protect them, shouldn't all the extra Higgs
doublets ``naturally" be superheavy? In other words, wouldn't a
multiplicity of Higgs doublets make the ``gauge hierarchy problem"
much worse, since there are now many such fields whose masses have
to be tuned? The answer is that family symmetry protects the masses
of the extra Higgs doublets and there is no extra tuning. We assume
that the mass-squared of the Standard Model Higgs field (the
lightest Higgs field in the $SO(4)_F$ 9-plet) is set
``anthropically". Under reasonable assumptions this means that it
must be negative and have magnitude of order (100 GeV)$^2$
\cite{abds, bk}. The other Higgs fields in the 9-plet have masses
that are tied to that of the Standard Model Higgs field by the
$SO(4)_F$ family symmetry. Their masses are therefore of order the
scale of $SO(4)_F$ breaking. This breaking is assumed to be
dynamical, and therefore can occur without fine-tuning at a low
enough scale to produce observable effects.

In a previous paper \cite{enp}, one of us proposed a much more
ambitious version of this model, in which unification of the
Standard Model gauge couplings was achieved through the group $SU(3)
\times SU(3) \times SU(3) \times Z_3$. This led to a much more
involved model. Here, by staying with the Standard Model gauge group
$G_{SM} = SU(3)_c \times SU(2)_L \times U(1)_Y$ we have a model that
is considerably simpler and easier to analyze.
\section{The Model}

\vspace{0.2cm}
The model has the gauge group $G_{SM} \times SO(4)_F \times SU(N)_{DSB}$,
where $SU(N)_{DSB}$ is a confining group that plays the role of
dynamically breaking the family group $SO(4)_F$. The field content
is shown in Table I:

\vspace{0.5cm}

\noindent {\large\bf Table I:} The field content of the model. $F$
stands for the $G_{SM}$ ``family" representation $(3,2, \frac{1}{6})
+ (\overline{3}, 1, -\frac{2}{3}) + (\overline{3}, 1, \frac{1}{3}) +
(1, 2, -\frac{1}{2}) + (1,1, 1)$; and $H$ for the $G_{SM}$ Higgs
representation $(1,2,-\frac{1}{2})$.

\newpage
\begin{tabular}{|l|l|l|}
\hline {\bf Field} & $G_{SM} \times SO(4)_F$ & {\bf Symbol}  \\
& $\times SU(N)_{DSB}$ & \\ \hline 4 families
& $(F, 4, 1)$ & $\psi^i = Q^i, (u^c)^i, (d^c)^i, L^i, (\ell^c)^i$ \\
& & \\ mirror family & $(\overline{F}, 1, 1)$ & $\overline{\psi} =
\overline{Q}, \overline{u^c}, \overline{d^c}, \overline{L},
\overline{\ell^c}$ \\ & & \\ Higgs doublets & $(H, 9, 1)$ &
$\Phi^{(ij)}$ \\ & & \\ messenger scalar fields & $(1, 4, 1)$ &
$\eta^i_I$
\\ & & \\
DSB fermions & $(1, 4, N)$ & $\chi^i$ \\ & & \\
DSB fermions & $4 \times (1, 1, \overline{N})$ & $\overline{\chi}_a, \;\; a=1,.., 4$ \\
& &
\\
\hline
\end{tabular}

\vspace{0.5cm }

\noindent In Table I and throughout the paper, the $SO(4)_F$ indices
are denoted by latin letters $i,j,k$ and range from 1 to 4. The fact
that the Higgs fields are in a rank-2 symmetric tensor multiplet of
$SO(4)$ allows them to couple directly by a renormalizable Yukawa
term to the quarks and leptons, schematically as $Y (\psi^i \psi^j)
\Phi^{(ij)}$. Note that $SO(4)_F$ symmetry and the pattern of its
breaking controls the form of $\langle \Phi^{(ij)} \rangle$ and thus
the form of the ``textures" of the quark and lepton mass matrices.
So we now consider how $SO(4)_F$ is broken and how this breaking is
communicated to the Standard Model fields.

The dynamical symmetry breaking is done by a $\langle
\overline{\chi}_a \chi^i \rangle$ condensate, where as shown in
Table I the $\chi^i$ are $N$'s of $SU(N)_{DSB}$ in a 4 of $SO(4)_F$
and the $\overline{\chi}_a$ are four $\overline{N}$'s of
$SU(N)_{DSB}$ that are singlets of $SO(4)_F$ with the subscript $a$
being just a label that distinguishes them. Since renormalizable
couplings of the $\chi, \overline{\chi}$ fields to the Standard
Model fields are forbidden by the gauge symmetries of the model, as
is easily seen, the Standard Model fields can only learn of the
breaking of the family group $SO(4)_F$ through ``messenger fields",
which are the $\eta^i_I$ shown in Table I. These are real scalars
that are vectors under $SO(4)_F$ and singlets under the other
groups. There are several such messenger multiplets, which are
distinguished by a capital latin subscript.

The $SO(4)_F$-breaking condensate $\langle \overline{\chi}_a \chi^i
\rangle$ generates vacuum expectation values (VEV) for the messenger
fields through the terms

\begin{equation}
f_{aI} \langle \overline{\chi}_a \chi^i \rangle \eta^i_I +
\frac{1}{2} M^2_{IJ} \eta^i_I \eta^i_J,
\end{equation}

\noindent where here and throughout we always sum over repeated
indices of any type. These terms give $\langle \eta^i_I \rangle = -
M^{-2}_{IK} f_{aK} \langle \overline{\chi}_a \chi^i \rangle$. If the
scale of the $\langle \overline{\chi} \chi \rangle$ condensate is
called $\Lambda^3$, and the mass of the messenger fields $\eta$ is
assumed to be superheavy (near the Planck scale), then the messenger
VEVs are typically of order $\Lambda^3/M^2_{P \ell}$. Since the
scale $\Lambda$ is set by dynamical symmetry breaking, it can
naturally be of any magnitude, depending on the $SU(N)_{DSB}$ gauge
coupling. Thus the VEVs of the messenger fields can be quite near
the weak scale in a ``technically natural" way. If we suppose that
the VEVs of the messenger fields are in the 10 to 1000 TeV range, as
will be assumed later, then $\Lambda$ is of order $10^{14}$ GeV.
This is the scale at which the local $SO(4)_F$ symmetry is broken,
and thus the mass scale of the $SO(4)_F$ gauge bosons, which are
consequently far too heavy to affect low-energy physics. And since
the messenger fields are superheavy, their exchange is also
irrelevant to low-energy physics. The VEVs of the messenger fields,
by contrast, can be small enough to produce significant effects at
low energy, and in particular to split the 9-plet of Higgs fields
and determine the pattern of quark and lepton masses. Note that
since the matrices $M^2_{IJ}$ and $f_{aI}$ in Eq. (1) are arbitrary
parameters, they can have a non-trivial and perhaps hierarchical
form, and therefore so can the VEVs of the messenger fields.

There are two types of renormalizable couplings of the messenger
fields to the Standard Model fields. They couple directly to the
fermions through terms that are schematically of the form $y_I
(\psi^i \overline{\psi}) \eta^i_I$. Such terms, which will be
discussed in more detail later, have the effect of ``mating" the
mirror family with one of the four families to give them a large
mass, leaving three light families.

The messenger fields also couple directly to the Higgs doublets
through a renormalizable term of the form

\begin{equation}
{\cal L}_{\Delta M^2_{\Phi}} = \frac{1}{2} \lambda_{KI} \Phi^{(ij)
\dag} \Phi^{(jk)} \eta^k_K \eta^i_I.
\end{equation}

\noindent Defining what we shall call the ``master matrix" $m^2$ by

\begin{equation}
(m^2)^{ij} \equiv  \lambda_{IJ} \langle \eta^i_I \rangle \langle
\eta^j_J \rangle,
\end{equation}

\noindent we can write the mass terms of the nine Higgs doublets as

\begin{equation}
\begin{array}{cc}
{\cal L}_{M^2_{\Phi}} & = - \frac{1}{2} M^2 \Phi^{(ij) \dag}
\Phi^{(ji)} - (m^2)^{ki} \Phi^{(ij) \dag} \Phi^{(jk)} \\ & \\
& = - \frac{1}{2} M^2 Tr[\Phi^{\dag} \Phi] -  Tr[m^2 \Phi^{\dag}
\Phi].
\end{array}
\end{equation}

\noindent The parameter $M^2$ in Eq. (4) is the overall
$SO(4)_F$-invariant mass of the Higgs 9-plet. The matrix $m^2$  in
Eq. (4) gives the splittings within the 9-plet. As a result of these
splittings, one linear combination of the $\Phi^{(ij)}$ is lighter
than the rest. It is assumed that anthropic tuning of the parameter
$M^2$ causes the mass-squared of this lightest doublet to be
negative and of order (100 GeV)$^2$, meaning that it is the Standard
Model Higgs field. (In other words, $M^2$ varies among domains or
subuniverses of the universe, so that there exist domains in which
the mass-squared of the lightest doublet has the value required for
life to be possible.) Let the Standard Model Higgs doublet be the
following linear combination: $\Phi_{SM} = \frac{1}{2}
\Sigma_{ij}a_{ij}\Phi^{(ij)}$, with $\Sigma_{ij} |a_{ij}|^2 = 2$,
where $a_{ij}$ (like $\Phi^{(ij)}$) is a symmetric traceless matrix.
It then follows that $\langle \Phi^{(ij)} \rangle = a_{ij} \langle
\Phi_{SM} \rangle = a_{ij} v/\sqrt{2}$. This directly gives a
non-trivial ``texture" for the mass matrices of the four families of
quarks and leptons, through the Yukawa terms of the form $Y (\psi^i
\psi^j) \Phi^{(ij)}$. One sees immediately, however, that it gives a
texture of exactly the same form ($\propto a_{ij}$) for the mass
matrices of the up quarks, down quarks, and charged leptons of the
four families. However, there are also the mass terms of the form
$y_I (\psi^i \overline{\psi}) \eta^i_I$ that couple the four
families to the mirror family. Since, as we shall now see, these
terms can be different for the up quarks, down quarks and charged
leptons, a realistic spectrum for the three light families of quarks
and leptons can result.

The quark and lepton Yukawa terms given schematically above have the
actual forms

\begin{equation}
\begin{array}{ll}
{\cal L}_{Yuk} & = {\cal L}_{4 \times 4} + {\cal L}_{4 \times 1}
\\ & \\
{\cal L}_{4 \times 4} & = Y_u \Phi^{(ij)*}(u^i u^{cj}) + Y_d
\Phi^{(ij)}(d^i d^{cj}) + Y_{\ell} \Phi^{(ij)}(\ell^i \ell^{cj}) \\ & \\
{\cal L}_{4 \times 1} & = y_Q^I \eta^i_I (u^i \overline{u} + d^i
\overline{d}) + y_u^I \eta^i_I (u^{ci} \overline{u^c}) + y_d^I
\eta^i_I (d^{ci} \overline{d^c}) \\ & \\ & + y_L^I \eta^i_I (\ell^i
\overline{\ell}) + y_{\ell}^I \eta^i_I (\ell^{ci}
\overline{\ell^c}).
\end{array}
\end{equation}

\noindent ${\cal L}_{4 \times 4}$ contains the Yukawa couplings of
the four families to each other, and ${\cal L}_{4 \times 1}$
contains the Yukawa couplings of the four families to the mirror
family. In order to express the mass terms coming from ${\cal L}_{4
\times 1}$ more compactly, it is convenient to define the following
vectors in the $SO(4)_F$ family space:

\begin{equation}
X_f^i \equiv \sum_I y_f^I \langle \eta^i_I \rangle/m, \;\;\; f=Q, u,
d, L, \ell.
\end{equation}

\noindent where $m \equiv Y_u v/\sqrt{2}$. Then the fermion mass
matrices have the forms
\begin{equation}
\begin{array}{l}
{\cal L}_{M,up} = Y_u \frac{v}{\sqrt{2}} \left( u^1, u^2, u^3, u^4,
\overline{u^c} \right) \left( \begin{array}{cccc|c} & & & &  X_Q^1 \\
& a_{ij} & & & X_Q^2 \\
& & & & X_Q^3 \\
& & & & X_Q^4 \\ \hline X_u^1 & X_u^2 & X_u^3 & X_u^4 & 0
\end{array} \right) \left( \begin{array}{c} u^{c1} \\
u^{c2} \\ u^{c3} \\ u^{c4} \\ \overline{u}
\end{array} \right), \\ \\
{\cal L}_{M,down} = Y_u \frac{v}{\sqrt{2}} \left( d^1, d^2, d^3,
d^4,
\overline{d^c} \right) \left( \begin{array}{cccc|c} & & & &  X_Q^1 \\
& ra_{ij} & & & X_Q^2 \\
& & & & X_Q^3 \\
& & & & X_Q^4 \\ \hline X_d^1 & X_d^2 & X_d^3 & X_d^4 & 0
\end{array} \right) \left( \begin{array}{c} d^{c1} \\
d^{c2} \\ d^{c3} \\ d^{c4} \\ \overline{d}
\end{array} \right), \\ \\
{\cal L}_{M,lepton} = Y_u \frac{v}{\sqrt{2}} \left( \ell^1, \ell^2,
\ell^3, \ell^4,
\overline{\ell^c} \right) \left( \begin{array}{cccc|c} & & & &  X_L^1 \\
& sa_{ij} & & & X_L^2 \\
& & & & X_L^3 \\
& & & & X_L^4 \\ \hline X_{\ell}^1 & X_{\ell}^2 & X_{\ell}^3 &
X_{\ell}^4 & 0
\end{array} \right) \left( \begin{array}{c} \ell^{c1} \\
\ell^{c2} \\ \ell^{c3} \\ \ell^{c4} \\ \overline{\ell}
\end{array} \right),
\end{array}
\end{equation}

\noindent where $r \equiv Y_d/Y_u$, $s \equiv Y_{\ell}/Y_u$. Note
that the elements in the $1 \times 4$ and $4 \times 1$ blocks of
these matrices are very large ($O(\langle \eta^i_I \rangle )$)
compared to the elements in the $4 \times 4$ blocks, which are
$O(\Phi^{(ij)})$, i.e. the weak scale or smaller. All these matrices
can be brought by change of bases to the general form

\begin{equation}
\left( \begin{array}{cccc|c} & & & & 0 \\
& A_{ij} & & & 0 \\
& & & & 0 \\
& & & & B \\ \hline 0 & 0 & 0 & C & 0
\end{array} \right).
\end{equation}

\noindent When this is done, one sees that the fermions of the
fourth family (in this basis) obtain very large Dirac masses with
the fermions of the mirror family, while the first three families
remain light.  The effective $3 \times 3$ mass matrix of the light
families is then just given by the first three rows and columns of
what we call $A_{ij}$ in Eq. (8) (with corrections that are $O(v/
\langle \eta \rangle)$) or smaller and thus utterly negligible). One
sees from this that the {\it magnitudes} of the ``vectors" $X_f^i$,
in the $1 \times 4$ and $4 \times 1$ blocks of the mass matrices in
Eq. (7) do not affect the spectrum of the light three families, only
their {\it directions} do.

The three mass matrices in Eq. (7) depend on several groups of
parameters. (a) $r$, $s$, which are just ratios of Yukawa couplings
($r \equiv Y_d/Y_u$, $s \equiv Y_{\ell}/Y_u$). (b) $a_{ij}$, which
is just the direction of the VEV of $\Phi^{(ij)}$ in $SO(4)_F$
space, and is determined by the mass matrix of the $\Phi^{(ij)}$,
which in turn is controlled by the ``master matrix" $m^2$ defined in
Eq. (3). And (c) the ``vectors" defined in Eq. (6). Most of the
parameters are in this last category.  These five vectors could be
independent of each other, in which case the number of parameters
would be too large to have a predictive model.

There are a number of ways in which the five vectors could be
related to each other, thus reducing the number of free parameters.
One is through unification of the Standard Model gauge group in a
larger group. This was the approach discussed in \cite{enp}, where
$G_{SM}$ was embedded in the ``trinification group" $SU(3) \times
SU(3) \times SU(3)$. Such unification symmetries relate quarks to
leptons and thus relate some of these vectors to each other. As can
be seen from \cite{enp}, however, there are significant costs to
such unification. It makes models considerably more involved.

Another possibility is that a small number of messenger fields give
the dominant contributions to the vectors of Eq. (6). To take an
extreme example, if only one messenger field, say $\eta_1^i$,
contributed, then the sums $\Sigma_I y_f^I \langle \eta_I^i \rangle$
in Eq. (6) would collapse to single terms proportional to $\langle
\eta_1^i \rangle$, and all the vectors would be parallel. This is
too extreme, however, because it would mean that the effective $3
\times 3$ mass matrices of the up quarks, down, quarks, and charged
leptons of the three light families would all be of the same form,
which is unrealistic.

An interesting possibility, which we will discuss briefly later, is
that all the vectors in Eq. (6) get their dominant contribution from
{\it two} of the messenger fields. Then the five vectors defined in
Eq. (6) would all lie in a two dimensional subspace.  The number of
parameters would thereby be reduced so much that the model would be
very predictive -- as predictive as the version of the model we
discuss below.

In this paper we follow a somewhat different path. We assume that
certain of the vectors (but not all of them) are dominated by a
single messenger field VEV and therefore parallel.  We will consider
two cases for illustration, which we will call ``Case A" and ``Case
B". In case A, the vectors $X_u$ and $X_Q$ are assumed parallel. In
case B, the vectors $X_d$ and $X_Q$ are assumed parallel. We will
only explicitly work out the quark sector couplings (the charged
lepton sector is quite similar, as will be seen), so we make no
assumption about the vectors $X_L$ and $X_{\ell}$ here.

\section{Fitting the quark spectrum in Case A}

We make the further assumption (to be justified later when we
discuss the spectrum of Higgs doublet masses) that the matrix
$a_{ij}$ is real. The forms of the mass matrices given in Eq. (7)
can then be simplified by a choice of $SO(4)_F$ basis. One can do an
$SO(4)_F$ transformation that makes the vectors $X_Q^i$ and $X_u^i$,
which are parallel in Case A, point in the 4 direction, i.e. have
the forms $(0,0,0,X_Q)$ and $(0,0,0,X_u)$. (This can be done with a
real orthogonal transformation, because these vectors are assumed
proportional to a single messenger field VEV, and each messenger
field is a {\it real} $SO(4)_F$ vector field. Moreover, because
$a_{ij}$ is real, an $SO(4)_F$ transformation preserves its
character as a {\it traceless} symmetric matrix. ) One can follow
this by an $SO(3)$ transformation involving only the indices
$i=1,2,3$ (which thus preserves the special forms of $X_Q^i$ and
$X_u$) that diagonalizes the upper-left $3 \times 3$ block of the
mass matrices. In the resulting basis the mass matrices have the
forms

\begin{equation}
\begin{array}{l}
M_{up} = \left( \begin{array}{cccc|c} c & 0 & 0 & d & 0 \\
0 & b & 0 & e & 0 \\
0 & 0 & a & f & 0 \\
d & e & f & -\Sigma & X_Q \\ \hline 0 \; & 0 \; & 0 \; & X_u & 0
\end{array} \right) m, \\ \\
M_{down} = r \left( \begin{array}{cccc|c}
c & 0 & 0 & d & 0 \\
0 & b & 0 & e & 0 \\
0 & 0 & a & f & 0 \\
d & e & f & -\Sigma & \frac{1}{r} X_Q \\ \hline \frac{1}{r} X_d^1 &
\frac{1}{r} X_d^2 & \frac{1}{r} X_d^3 & \frac{1}{r} X_d^4 & 0
\end{array} \right) m, \\ \\
M_{lep} = s \left( \begin{array}{cccc|c} c & 0 & 0 & d & \frac{1}{s} X_L^1 \\
0 & b & 0 & e & \frac{1}{s} X_L^2 \\
0 & 0 & a & f & \frac{1}{s} X_L^3 \\
d & e & f & -\Sigma & \frac{1}{s} X_L^4 \\ \hline \frac{1}{s}
X_{\ell}^1 & \frac{1}{s} X_{\ell}^2 & \frac{1}{s} X_{\ell}^3 &
\frac{1}{s} X_{\ell}^4 & 0
\end{array} \right) m,
\end{array}
\end{equation}

\noindent where $\Sigma \equiv a + b + c$. Note that because we have
obtained this form by a real orthogonal transformation, and because
we are assuming that $a_{ij}$ is real, the parameters $a,b,c,d,e,f$
in these matrices are real.

Since $X_Q$ and $X_u$ (which are of order $\langle \eta^i_I
\rangle$) are several orders of magnitude larger than the elements
$a, b, c, d, e, f$, it is easily seen that the three light families
of up quarks (namely $u,c,t$) correspond almost exactly to the first
three rows and columns of $M_{up}$ in Eq. (9). Thus, the effective
mass matrix for the three observed families of up quarks is given in
this basis simply by

\begin{equation}
\tilde{M}_{up} = \left( \begin{array}{ccc} c & 0 & 0 \\
0 & b & 0 \\ 0 & 0 & a \end{array} \right) m.
\end{equation}

\noindent Therefore, $c/b = m_u/m_c \ll 1$ and $b/a = m_c/m_t \ll
1$. Since it will turn out that $b,c,d,e,f$ are all small compared
to 1, and $a_{ij}$ is normalized so that $\Sigma_{ij} |a_{ij}|^2 =
2$, one has $a^2 \cong 1$. Without loss of generality we can take $a
\cong +1$, and $m \equiv Y_u v/\sqrt{2} \cong m_t$.

To find the effective mass matrix for the three light families of
down quarks, we must do a further change of basis of the $d^{ci}$ to
bring the complex vector $(X_d^1, X_d^2, X_d^3, X_d^4)$ to the form
$(0,0,0,X_d)$. This is done by multiplying $M_{down}$ from the right
by a unitary transformation of the form

\begin{equation}
\begin{array}{cl}
U & = \left( \begin{array}{cccc} c_{\alpha} & s_{\alpha}^* & 0 & 0 \\
-s_{\alpha} & c_{\alpha} & 0 & 0 \\ 0 & 0 & 1 & 0 \\ 0 & 0 & 0 & 1
\\ \end{array} \right)
\left( \begin{array}{cccc} 1 & 0 & 0 & 0 \\
0 & c_{\beta} & s_{\beta}^* & 0  \\ 0 & -s_{\beta} & c_{\beta} & 0 \\
0 & 0 & 0 & 1
\\ \end{array} \right)
\left( \begin{array}{cccc} 1 & 0 & 0 & 0 \\
0 & 1 & 0 & 0 \\ 0 & 0 & c_{\gamma} & s_{\gamma} \\ 0 & 0 &
-s_{\gamma} & c_{\gamma}^*
\\ \end{array} \right) \\ & \\
& = \left( \begin{array}{cccc} c_{\alpha} & c_{\beta} s_{\alpha}^* &
c_{\gamma} s_{\beta}^* s_{\alpha}^* & s_{\gamma} s_{\beta}^* s_{\alpha}^* \\
-s_{\alpha} & c_{\beta} c_{\alpha} &
c_{\gamma} s_{\beta}^* c_{\alpha} & s_{\gamma} s_{\beta}^* c_{\alpha} \\
0 & -s_{\beta} & c_{\gamma} c_{\beta} & s_{\gamma} c_{\beta} \\
0 & 0 & -s_{\gamma} & c_{\gamma}^*
\end{array}
\right) \end{array}
\end{equation}

\noindent Where the angles $s_{\alpha}$, $s_{\beta}$ and
$s_{\gamma}$ are in general complex. It turns out that to get a
realistic fit to the quark masses, one needs to assume that $c,d \ll
b,e \ll f < a \cong 1$, and that $|s_{\alpha}|$, $|s_{\beta}|$, and
$|c_{\gamma}|$ are small compared to 1. This allows us to write
$M_{down}$ in the new basis as

\begin{equation}
M_{down} \cong r \left( \begin{array}{cccc|c} 0 & 0 & -d &
c_{\gamma}^* & 0
\\ -s_{\alpha} b & b & -e & s_{\beta}^* +c_{\gamma}^* e & 0 \\
0 & -s_{\beta}  & c_{\gamma}  - f & 1 + c_{\gamma}^* f & 0 \\
d - s_{\gamma} e & e - s_{\beta} f & c_{\gamma} f + 1 & f -
c_{\gamma}^*  & \frac{1}{r} X_Q \\ \hline 0 & 0 & 0 & \frac{1}{r}
X_d & 0
\end{array} \right) m
\end{equation}

\noindent From this one can read off that the effective $3 \times 3$
mass matrix of the three light families of down quarks is simply

\begin{equation}
\tilde{M}_{down} \cong r \left( \begin{array}{ccc} 0 & 0 & -d
\\ -s_{\alpha} b & b & -e \\
0 & -s_{\beta}  & c_{\gamma}  - f
\end{array} \right) m
\end{equation}

\noindent The parameter $s_{\alpha}$ can be made real by redefining
the phase of $d^{c1}$ in this basis. The parameter $s_{\beta}$ can
be made real by redefining the phase of $u^3$ in this basis. (These
phase redefinitions do not affect the fitting of known quantities,
but do affect the phases of the Yukawa couplings of the ``extra"
scalar doublets, which are therefore undetermined by just fitting
the known quark masses and mixing angles.) Calling $c_{\gamma} -f
\equiv F e^{i \phi}$ and remembering that we have normalized $a$ to
be 1, the quark mass matrices can be written

\begin{equation}
\tilde{M}_{up} = \left( \begin{array}{ccc} c & 0 & 0 \\
0 & b & 0 \\ 0 & 0 & 1 \end{array} \right) m, \;\; \tilde{M}_{down}
\cong r \left( \begin{array}{ccc} 0 & 0 & -d
\\ -s_{\alpha} b & b & -e \\
0 & -s_{\beta}  & F e^{i\phi}
\end{array} \right) m.
\end{equation}

\noindent These depend on nine real parameters ($r, m, b, c, d, e,
F, s_{\alpha}, s_{\beta}$) and one phase ($e^{i \phi}$).  This is
just the right number of parameters to fit the six quark masses,
three CKM angles and the CKM phase. The results of the fit are given
in Table II.

\vspace{0.5cm}

\noindent {\large\bf Table II:} Parameter values in Case A of the
model that reproduce the known quark masses and CKM mixing matrix.

\vspace{0.2cm}

\begin{displaymath}
\begin{array}{l|l}
\hline {\bf parameter} &  {\bf value}  \\ \hline a & 1.0 \\
b & 3.6 \times 10^{-3} \\
c & 7.4 \times 10^{-6} \\
d & 4.7 \times 10^{-4} \\
e & 2.2 \times 10^{-3} \\
F & 5.7 \times 10^{-2} \\
\phi & 0.98 \\
\sin \alpha & 0.105 \\
\sin \beta & 0.076 \\
r & 0.177 \\ \hline
\end{array}
\end{displaymath}

\vspace{0.5cm}

\noindent Note that the parameter $f$ is not determined, and the
parameter $c_{\gamma}$ is given by $c_{\gamma} = F e^{i \phi} + f$.
These numbers determine (except for the parameter $f$) the $4 \times
4$ mass matrix of the four families in the basis of Eq. (9):

\begin{equation}
Y_u \langle \Phi^{(ij)} \rangle = Y_u \left( \frac{v}{\sqrt{2}}
\right) a_{ij} = Y_u \frac{v}{\sqrt{2}} \left(
\begin{array}{cccc} c & 0 & 0 & d
\\ 0 & b & 0 & e \\ 0 & 0 & a & f \\ d & e & f & - \Sigma
\end{array} \right).
\end{equation}

\noindent In the next section, we will use this information to
determine the spectrum of the scalars $\Phi^{(ij)}$. This is
possible because the matrix $a_{ij}$ is enough to determine the
master matrix $(m^2)^{ij}$ (if that is assumed real).

\section{The scalar spectrum in Case A}

The masses of the Higgs doublets $\Phi^{(ij)}$ are controlled by the
``master matrix" $m^2$ defined in Eq. (3). (There are also
contributions to the mass-squared of the ``extra" scalar doublets
that come from the the quartic self-couplings of $\Phi^{(ij)}$ once
$\Phi_{SM}$ gets a VEV, but these are negligible if, as will turn
out to be the case, the masses of the ``extra" doublets are much
larger than the mass of the Standard Model Higgs.) From Eq. (3), one
easily sees that $m^2$ is hermitian. (In that equation the coupling
matrix $\lambda_{IJ}$ is in general complex and hermitian, whereas
the VEVs $\langle \eta^i_I \rangle$ are real.) To obtain a realistic
hierarchy among the quark and lepton masses, it turns out that $m^2$
must be very hierarchical, as will be seen. In simple cases where
$m^2$ is hierarchical, it also tends to be approximately real. (To
take an extreme case, suppose, that one of the $\eta_I^i$, say
$\eta_1^i$, gave the largest contribution to $m^2$. Then $(m^2)^{ij}
\cong \lambda_{11} \langle \eta^i_1 \rangle \langle \eta_1^j
\rangle$, which is rank-1, and thus hierarchical, and also
manifestly real.) We therefore make the approximation that $m^2$ is
real, since this greatly simplifies the analysis of the model. (It
is also possible to imagine that the master matrix arises primarily
from the VEVs of other messenger fields $\eta^{(ij)}$ that are real
9-plets of $SO(4)_F$. Those contributions would be {\it exactly}
real.)

If $m^2$ is taken to be real, then it is also symmetric, and it can
be diagonalized by an $SO(4)_F$ rotation, i.e. by a choice of
$SO(4)_F$ basis, which we will call the ``scalar-mass basis". Since
the terms in Eq. (4) can be written $Tr[(\frac{1}{2} M^2 I + m^2)
\Phi^{\dag} \Phi]$, it is clear that without loss of generality one
can make one of the diagonal elements of $m^2$ vanish by shifting
the parameter $M^2$. Thus $m^2$ can be taken (in the ``scalar-mass
basis") to be of the form

\begin{equation}
m^2 = \left( \begin{array}{cccc} 1 & 0 & 0 & 0 \\
0 & \epsilon & 0 & 0 \\ 0 & 0 & \delta & 0 \\
0 & 0 & 0 & 0 \end{array} \right) m_0^2.
\end{equation}

\noindent We assume that $\delta \ll \epsilon \ll 1$, which will
lead directly to a hierarchy in the quark and lepton mass matrices,
as will be seen. Writing the 9-plet of Higgs fields as

\begin{equation}
\Phi^{(ij)} = \left(  \begin{array}{cccc} \frac{3
\overline{\Phi}_{11}}{\sqrt{6}}
& \Phi_{12} & \Phi_{13} & \Phi_{14} \\
\Phi_{12} & \frac{2 \sqrt{2} \; \overline{\Phi}_{22} -
\overline{\Phi}_{11}}{\sqrt{6}}
& \Phi_{23} & \Phi_{24} \\
\Phi_{13} & \Phi_{23} &  \frac{\sqrt{6} \; \overline{\Phi}_{33} -
\sqrt{2} \; \overline{\Phi}_{22} - \overline{\Phi}_{11}}{\sqrt{6}}
& \Phi_{34} \\
\Phi_{14} & \Phi_{24} & \Phi_{34} & \frac{- \sqrt{6} \;
\overline{\Phi}_{33} - \sqrt{2} \; \overline{\Phi}_{22} -
\overline{\Phi}_{11}}{\sqrt{6}}
\end{array} \right)
\end{equation}

\noindent and substituting Eqs. (16) and (17) into Eq. (4), one
finds the spectrum given in Table III:


\noindent {\large\bf Table III:} The mass spectrum of the 9-plet of
Higgs doublets.

\vspace{0.2cm}

\begin{tabular}{lll}
{\bf Field} & $\;\;\; $(mass)$^2$ & after tuning SM Higgs  \\
\hline $\overline{\Phi}^{(11) \prime}$ & $\cong M^2 + \frac{3}{2}
m_0^2$ &
$\cong \frac{3}{2} m_0^2$ \\
$\Phi^{(12)}$ & $\;\;\; M^2 + (1+ \epsilon) m_0^2$ & $\cong (1 +
\epsilon) m_0^2$
\\
$\Phi^{(13)}$ & $\;\;\; M^2 + (1 + \delta) m_0^2$ & $\cong m_0^2$
\\
$\Phi^{(14)}$ & $\;\;\; M^2 + m_0^2$ & $\cong (1 - \delta) m_0^2$
\\ & & \\
$\overline{\Phi}^{(22)\prime}$ & $\cong M^2 + \frac{1}{3} \epsilon
m_0^2$ & $\cong \frac{1}{3} \epsilon m_0^2$ \\
$\Phi^{(23)}$ & $\;\;\; M^2 + (\epsilon + \delta)
m_0^2$ & $\cong \epsilon m_0^2$ \\
$\Phi^{(24)}$ & $\;\;\; M^2 + \epsilon m_0^2$ &
$\cong (\epsilon - \delta) m_0^2$ \\ & & \\
$\Phi^{(34)}$ & $\;\;\; M^2 + \delta m_0^2$ & $\cong
\frac{\delta^2}{4 \epsilon} m_0^2$ \\
$\overline{\Phi}^{(33)\prime} (\equiv \Phi_{SM})$ & $\cong M^2 +
(\delta - \frac{\delta^2}{4 \epsilon}) m_0^2$ & $\equiv - \mu^2$
\end{tabular}

\vspace{0.5cm}

\noindent Note that the Higgs fields in the first row/column
($\Phi^{(1i)}$) get contributions of order $m_0^2$ from the master
matrix; those in the second (but not first) row/column
($\Phi^{(2i)}, i \neq 1$) get contributions of order $\epsilon
m_0^2$, and the remaining ones get contributions of order $\delta
m_0^2$, as an inspection of Eqs. (4) and (15) would suggest. The
fields denoted $\overline{\Phi}^{(ii) \prime}$ are linear
combinations of the fields denoted $\overline{\Phi}^{(ii)}$ in Eq.
(17). The lightest of the Higgs doublets, which is the Standard
Model Higgs doublet, turns out to be the linear combination

\begin{equation}
\Phi_{SM} = \overline{\Phi}^{(33) \prime} \cong
\overline{\Phi}^{(33)} + \frac{\sqrt{3}}{4} \frac{\delta}{\epsilon}
\overline{\Phi}^{(22)} + \frac{5}{6 \sqrt{6}} \delta
\overline{\Phi}^{(11)}.
\end{equation}

\noindent One sees, then, that the Standard Model Higgs doublet has
diagonal Yukawa couplings in the ``scalar-mass basis".  The
mass-squared of the Standard Model Higgs doublet is fine-tuned
(presumably ``anthropically") to be $- \mu^2$, where $\mu \sim 100$
GeV. This gives $M^2 \cong - (\delta - \frac{\delta^2}{4 \epsilon})
m_0^2 - \mu^2$. Substituting this into the mass-squared of the other
Higgs fields in the 9-plet gives the results in the last column of
Table III.

The next lightest Higgs doublet is $\Phi^{(34)}$. We will call this
$\Phi_{LED}$, where LED stands for ``lightest extra doublet". From
Table III, one sees that the mass of $\Phi_{LED}$ is
$\frac{\delta}{2 \epsilon}$ times that of the next lightest Higgs
doublets $\Phi^{(23)}$ and $\Phi^{(24)}$.  Shortly, we will see that
this is $3.6 \times 10^{-3}$. Thus, it turns out that
flavor-violating effects are dominated by the exchange of
$\Phi_{LED}$. In the scalar-mass basis, $\Phi_{LED} = \Phi^{(34)}$
couples very simply to the quarks and leptons: it only couples the
third to the fourth family, with strength 1 for the up quarks, $r$
for the down quarks, and $s$ for the charged leptons.

From Eqs. (18) one sees that $\langle \overline{\Phi}^{(11)} \rangle
= \left( \frac{5 \delta}{6\sqrt{6}} \right) v/\sqrt{2}$, $\langle
\overline{\Phi}^{(22)} \rangle = \left( \frac{\sqrt{3} \delta}{4
\epsilon} \right) v/\sqrt{2}$, and $\langle \overline{\Phi}^{(33)}
\rangle = v/\sqrt{2}$. Substituting this into Eq. (17), one finds
that the matrix $a_{ij}$ that appears in the mass matrices given in
Eq. (7) is just given in the scalar-mass basis by

\begin{equation}
\langle \Phi^{(ij)} \rangle = a_{ij}v/ \sqrt{2} \cong \left(
\begin{array}{cccc}
\frac{5}{12} \delta & 0 & 0 & 0 \\
0 & \frac{1}{2} (\delta/\epsilon) & 0 & 0 \\
0 & 0 & 1 & 0 \\ 0 & 0 & 0 & -1 \end{array} \right) v/ \sqrt{2}.
\end{equation}

\noindent This is related to the form of $a_{ij}$ in Eqs. (9) and
(15) by a change of basis of the fermions. Indeed, since the
parameters that appear in Eq. (15) are given in Table II (except for
$f$), one simply diagonalizes the form in Eq. (15) to determine the
parameters $\epsilon$ and $\delta$ in Eq. (19). In this way, one
finds that $\epsilon \cong 2.5 \times 10^{-3}$, $\delta \cong 1.8
\times 10^{-5}$, and $\delta/\epsilon \cong 7.2 \times 10^{-3}$

Since the transformation between these two bases is known (in terms
of one unknown, namely $f$), one can determine the Yukawa couplings
of $\Phi_{LED}$, and indeed all the other extra Higgs doublets, in
the basis of Eqs. (9) and (15). The basis of Eqs. (9) and (15) is in
fact the physical basis of the up quarks $u$, $c$ and $t$, as
explained before Eq. (10). Thus we know how all nine of the Higgs
doublets couple to $u$, $c$, and $t$. To get to the physical basis
of the down quarks $d$, $s$, and $b$, one must do two further
changes of basis of the down quarks: first, that shown in Eq. (11),
the parameters of which are given in Table II (except for the phases
of $s_{\alpha}$ and $s_{\beta}$); and second, the change of basis
needed to diagonalize the matrix in Eq. (13), which are completely
determined from Table II.

In other words, one is in a position to compute the couplings of the
{\it all} nine of the the $\Phi^{(ij)}$ to {\it all} of the known
quarks in terms of the unknown parameter $f$ and the unknown phases
of $s_{\alpha}$ and $s_{\beta}$.

The results are given in Table IV, for $f = 0.05$ and the phases of
$s_{\alpha}$ and $s_{\beta}$ equal to zero. One gets similar results
for other values of these parameters.


\noindent {\large\bf Table IV:} Values for the Yukawa couplings of
the lightest extra Higgs doublet (LED) to the quarks, in Case A of
the model, with $f = 0.05$, and $s_{\alpha} = s_{\beta} = 0$.

\vspace{0.2cm}

\begin{displaymath}
\begin{array}{c|l}
\hline {\bf Yukawa \; of \; LED} &  {\bf value}  \\ \hline
Y^u_{12} = Y^u_{21} & 2.1 \times 10^{-7} \\
Y^u_{13} = Y^u_{31} & -4.7 \times 10^{-4}  \\
Y^u_{23} = Y^u_{32} & 2.2 \times 10^{-3}  \\
Y^u_{11} & -1.1 \times 10^{-8} \\
Y^u_{22} & -9.7 \times 10^{-7} \\
Y^u_{33} & 0.12 \\
Y^d_{12} & (-0.37 -i 1.2) \times 10^{-3} \\
Y^d_{21} & (6.2 + i 8.5) \times 10^{-4} \\
Y^d_{13} & (2.8 + i 9.4) \times 10^{-4}  \\
Y^d_{31} & (1.5 - i 2.3) \times 10^{-2} \\
Y^d_{23} & (2.4 +i 0.34) \times 10^{-3}  \\
Y^d_{32} & (-0.77 + i 1.2) \times 10^{-1} \\
Y^d_{11} & (0.7 + i 2.3) \times 10^{-4} \\
Y^d_{22} & (3.2 + i 4.4) \times 10^{-3} \\
Y^d_{33} & (6.0 - i 8.9) \times 10^{-2} \\ \hline
\end{array}
\end{displaymath}

\vspace{0.5cm}

The analysis of the charged leptons is very similar to that of the
down quarks. There are a number of assumptions that could be made
about the vectors $X_L^i$ and $X_{\ell}^i$ in Eq. (9). Suppose, for
example, one assumed that $X_{\ell}^i$ is parallel to $X_Q^i$ and
$X_u^i$. Then the diagonalization of $M_{lep}$ proceeds in the same
way as the diagonalization of $M_{down}$ above, except that
$M_{lep}$ in Eq. (9) is multiplied on the {\it left} by a unitary
matrix $U^{\prime \dag}$, where $U'$ has the same form as $U$ in Eq.
(11) but with different angles $\alpha'$, $\beta'$, and $\gamma'$.
The phases of these parameters turn out not to affect the fitting of
the charged lepton masses significantly.  So there are four
additional parameters in the lepton sector ($s$, $\alpha'$,
$\beta'$, and $\gamma'$) available to fit the three masses $m_e$,
$m_{\mu}$, and $m_{\tau}$. Consequently, the Yukawa coupling
matrices of all 9 Higgs doublets to the charged leptons are
determined in terms of only a small number of additional unknown
parameters. Here we will only discuss the quark sector for purposes
of illustration.

The Yukawa couplings in Table IV allow us to write down the
coefficients flavor-changing four-fermion operators. The most
interesting involving the down-type quarks are given in Table V.


\noindent {\large\bf Table V:} The predicted coefficients of the
most important flavor-changing four-quark operators and the
resulting lower limits on the mass of the LED \cite{fcnc}.

\vspace{0.2cm}

\begin{displaymath}
\begin{array}{c|c|l}
\hline {\bf Operator} &  {\bf Coefficient} & {\bf Limit \;\; on \;\;} M_{LED} \\
\hline & & \\
c_{sd} (\overline{s_R} d_L)(\overline{s_L} d_R) &
|c_{sd}| = 1.33 \times 10^{-6}/M_{LED}^2 & \geq 14 {\rm TeV} \\
 & Im (c_{sd}) =  1.33 \times 10^{-6}
 \frac{\arg(c_{sd})}{M_{LED}^2}
 & \geq 230 {\rm TeV} [\arg(c_{sd})]^{1/2} \\
& & \\ c_{bs} (\overline{b_R} s_L)(\overline{b_L} s_R) & |c_{bs}| =
3.45 \times 10^{-4}/M_{LED}^2 & \geq 5.1 \; {\rm TeV} \\
& & \\
c_{bd} (\overline{b_R} d_L)(\overline{b_L} d_R) & |c_{bd}| = 2.7
\times 10^{-5}/M_{LED}^2 & \geq 6.9 {\rm TeV} \\ \hline
\end{array}
\end{displaymath}

\vspace{0.5cm}

In Table V, the limits on $M_{LED}$ are obtained from the limits on
the coefficients of flavor-changing operators given in \cite{fcnc}.
One sees from Table V that the contribution to $\epsilon_K$ from the
CP-violating part of the $(\overline{s_R} d_L)(\overline{s_L} d_R)$
operator gives an extremely severe constraint on the mass of the
lightest extra doublet in this model if the phase of $c_{sd}$ is
order one. If that phase happens to be very small, then $\delta m_K$
still constrains the LED mass to be greater than 14 TeV. (It should
be pointed out that these numbers turn out to be fairly insensitive
to the value of the unknown parameter $f$.)

These bounds are considerably tighter than one might have expected
for a flavor-changing Higgs if its Yukawa couplings were similar to
those of the Standard Model Higgs. These bounds are very sensitive
to the details of the model. We will now look at another version of
the model (Case B), since the comparison is instructive.

\section{Results for Case B}

The analysis of Case B is quite similar to that of Case A. In case B
the mass matrices in the same basis as Eq. (9) take the form

\begin{equation}
\begin{array}{l}
M_{up} = \left( \begin{array}{cccc|c} c & 0 & 0 & d & 0 \\
0 & b & 0 & e & 0 \\
0 & 0 & a & f & 0 \\
d & e & f & -\Sigma & X_Q \\ \hline X^1_u \; & X^2_u \; & X^3_u \; &
X^4_u & 0
\end{array} \right) m, \\ \\
M_{down} = r \left( \begin{array}{cccc|c}
c & 0 & 0 & d & 0 \\
0 & b & 0 & e & 0 \\
0 & 0 & a & f & 0 \\
d & e & f & -\Sigma & \frac{1}{r} X_Q \\ \hline 0 & 0 & 0 &
\frac{1}{r} X_d & 0
\end{array} \right) m.
\end{array}
\end{equation}

\noindent In this case, it is apparent that (neglecting terms of
order $v/\langle \eta \rangle$) the mass matrix of the observed down
quarks, $d$, $s$, $b$ is just given by the upper left $3 \times 3$
block of $M_{down}$, i.e.

\begin{equation}
\tilde{M}_{down} = r \left( \begin{array}{ccc} c & 0 & 0 \\
0 & b & 0 \\ 0 & 0 & a \end{array} \right) m,
\end{equation}

\noindent so that this is already in the physical basis of these
quarks. Thus, in case B, $b/a = m_s/m_b$ and $c/a = m_d/m_b$.

For the up-type quarks, however, one must make a further change of
basis for the $u^{ci}$ in order to bring the complex vector $(X_u^1,
X_u^2, X_u^3, X_u^4)$ to the form $(0,0,0, X_u)$. This involves
rotating the matrix $M_{up}$ in Eq. (20) from the right by a matrix
of the same form shown in Eq. (11). This gives for $M_{up}$ the same
form as $M_{down}$ has in Case A, shown in Eq. (12) (with $r =1$).
However, it turns out that the fit to the quark masses and mixing
angles implies that here $a <1$ and $f \cong 1$, and, unlike Case A,
the parameter $c$ is not negligible. Moreover, the angle $\beta$ is
large enough here that it is not a good approximation to set $\cos
\beta =1$, but it is a good approximation here, as in Case A, to set
$\cos \alpha = 1$ and $\sin \gamma = 1$.  With these approximations,
one has for the effective $3 \times 3$ mass matrix of the $u$, $c$,
and $t$ quarks

\begin{equation}
\tilde{M}_{up} \cong  \left( \begin{array}{ccc} c & c_{\beta}
s_{\alpha} c & -d
\\ -s_{\alpha}^* b & c_{\beta} b & -e \\
0 & -s_{\beta}^* a  & - f
\end{array} \right) m
\end{equation}

\noindent Fitting the quark masses and mixing angles leads to the
the parameter values given in Table VI.

\vspace{0.5cm}

\noindent {\large\bf Table VI:} The parameter values in Case B of
the model that reproduce the masses of the quarks and the CKM mixing
matrix.

\vspace{0.2cm}

\begin{displaymath}
\begin{array}{l|l}
\hline {\bf parameter} &  {\bf value}  \\ \hline a & 0.137 \\
b & 2.54 \times 10^{-3} \\
c & 1.27 \times 10^{-4} \\
d & 7.5 \times 10^{-3} \\
e & -0.04 \\
f & 1.0 \\
\sin \alpha & 0.24 e^{i 0.42} \\
\sin \beta & 0.08 e^{i 1.05} \\
r & 0.132 \\ \hline
\end{array}
\end{displaymath}

\noindent It turns out that $\cos \gamma$ is almost unconstrained,
since the quark masses and mixing angles are only very weakly
dependent on it. With the parameters given in Table VI, the Yukawa
couplings of all the Higgs doublets to all the quarks can be
straightforwardly computed in terms of $\cos \gamma$.

What distinguishes Cases A and B is that in case A the strongest
flavor-changing effects are in the down-quark sector, whereas for
Case B the strongest flavor-changing effects are in the up-quark
sector. What matters most in Case B, therefore, are the couplings of
$u$ to $c$, which give the operator $(\overline{c_L}
u_R)(\overline{c_R} u_L)$. The coefficient of this operator $c_{uc}$
 is somewhat insensitive to the value of $\cos \gamma$. With
$\cos \gamma = 0.1$, one finds $c_{uc} = 7.5 \times
10^{-5}/M_{LED}^2$. The current limit from $D-\overline{D}$ mixing
gives $M_{LED} \geq 36$ TeV. The limits from the $B_s$ and $B_d$
system turn out to be much weaker: they only constrain $M_{LED}$ to
be larger than about 1.4 TeV. The limit from the $\epsilon_K$
parameter is that $M_{LED} > 7$ TeV, for CP phases of order 1.

One sees, then, that in the two special cases of the model that we
have analyzed the lightest extra Higgs doublet has to be too heavy
to be seen at accelerators or to give significant flavor-changing
effects in rare processes. In Case A this because of the
$K-\overline{K}$ mixing limits and in Case B it is because of the
$D-\overline{D}$ mixing limits. However, Cases A and B do not
exhaust the possibilities of this model. For example, as noted near
the end of section 2, the assumption that all the ``vectors" in Eq.
(6) arise from just two messenger fields reduces the number of
parameters almost as much as in the two cases we have studied here.
It may be that this assumption or other assumptions or limits of the
model can allow the lightest extra Higgs doublet to be lighter than
in Cases A and B. Moreover, what has been studied here is only one
particular model that realizes the basic idea of putting multiple
Higgs doublets into a representation of a non-abelian flavor group.

\section{Conclusions}

The repetition of quark and lepton families has long suggested the
possibility of a non-abelian family symmetry \cite{family}. It is
quite natural, therefore, to consider the possibility that the Higgs
field of the Standard Model belongs to a multiplet of the same
family group. In the model we have presented as an example of this
idea, the family symmetry tightly constrains the forms of the quark
and lepton mass matrices. Nevertheless, it has been shown that the
observed fermion masses and mixing angles can be reproduced. The
family symmetry also severely constrains the Yukawa couplings of all
the ``extra" Higgs doublets; and it has been seen that after fitting
the known quark and lepton masses, the coefficients of all the
flavor-changing four-fermion operators that come from the exchange
of extra Higgs doublets are predicted in terms of only a few
parameters. It turns out that in the specific model we have studied,
the constraints from limits on flavor-changing in the
$K-\overline{K}$ and $D-\overline{D}$ systems require the lightest
extra Higgs doublet to have a mass of tens of TeV, which is too
heavy to lead to testable phenomenology in the near future. This
may, however, be a feature of the specific model we have studied
rather than an inevitable consequence of the general approach we are
proposing.

One of the interesting features of the approach being described in
this paper is that the spectrum of the Higgs fields is closely
connected to the spectrum of the quarks and leptons. The pattern of
couplings of the Standard Model Higgs to the quark families --- i.e.
the so-called ``textures" of the Yukawa matrices --- is determined
by which component within the ``family" of Higgs fields is the
lightest, i.e. is the Standard Model Higgs field. This is
determined, in turn, by the pattern of family-symmetry-breaking
within the Higgs family multiplet. Thus, both the spectrum of
fermion masses and the spectrum of Higgs boson masses is largely
determined by what we have called a ``master matrix".

A key feature of the present approach is that the breaking of the
family symmetry takes place dynamically in a sector of fields that
are Standard Model singlets and is communicated to the Standard
Model degrees of freedom by ``messenger" fields. If the messenger
sector is simple, then the pattern of masses of the Standard Model
fields, including the extra Higgs doublets, is highly constrained.

It would be interesting to see if other non-abelian family groups
and particular choices of family representations for the quarks,
leptons, and Higgs fields could lead to realistic models that
predict flavor-changing effects at observable levels.


\begin{thebibliography}{999}
\bibitem{family} T. Maehara and T. Yanagida, {\it Prog. Theor.
Phys.} {\bf 60}, 822 (1978); {\it ibid.} {\bf 61}, 1434 (1979); F.
Wilczek and A. Zee, {\it Phys. Rev. Lett.} {\bf 42}, 421 (1979);
R.N. Cahn and H. Harari, {\it Nucl. Phys.} {\bf B176}, 135 (1980);
O. Shanker, {\it Phys. Rev.} {\bf D23}, 1555 (1981); D.R.T. Jones,
G.L. Kane, {\it Nucl. Phys} {\bf B198}, 45 (1982); D.B. Reiss, {\it
Phys. Lett.} {\bf B115}, 217 (1982).
\bibitem{abds} V. Agrawal, S.M. Barr, J.F. Donoghue, and D. Seckel,
{\it Phys. Rev.} {\bf D57}, 5480 (1998); {\it ibid.} {\it Phys. Rev.
Lett.} {\bf 80} 1822 (1998).
\bibitem{bk} S.M. Barr and Almas Khan, {\it Phys. Rev.} {\bf
D76}, 045002 (2007).
\bibitem{enp} S.M. Barr, {\it Phys. Rev.} {\bf D82}, 055010
(2010).
\bibitem{fcnc} G. Isidori, Y. Nir, and G. Perez, {\it Annu. Rev.
Nucl. Part. Sci.} {\bf 60}, 355 (2010).
\end{thebibliography}
\end{document}